\documentclass[final,5p,fleqn]{elsarticle}

\usepackage{epsfig}
\usepackage{subfigure}
\usepackage[utf8]{inputenc}
\usepackage{float}
\usepackage{graphicx}
\usepackage{amsmath}
\usepackage{verbatim}
\usepackage{xcolor}
\usepackage{caption}
\usepackage{color}
\usepackage{soul}
\begin{document}

\begin{frontmatter}

\title{ 
Extraction of Kaon Partonic Distribution Functions from\\
Drell-Yan and $J/\psi$ Production Data} 

\author[a]{Claude Bourrely}
\author[b]{Franco Buccella}
\author[c]{Wen-Chen Chang}
\author[d]{Jen-Chieh Peng}

\address[a]{Aix Marseille Univ, Universite de Toulon, CNRS, CPT,
  Marseille, France}

\address[b]{INFN, Sezione di Roma I, Roma, Italy}

\address[c]{Institute of Physics, Academia Sinica, Taipei 11529, Taiwan} 

\address[d]{Department of Physics, University of Illinois at
  Urbana-Champaign, Urbana, Illinois 61801, USA}

\begin{abstract}
We present an analysis to extract kaon parton distribution 
functions (PDFs) using meson-induced Drell-Yan and 
quarkonium production
data. Starting from the statistical model first developed for determining
the partonic structure of spin-1/2 nucleon and later applied to the 
spin-0 pion, we have extended this approach to perform a global fit to existing 
kaon-induced Drell-Yan and $J/\psi$ production data.
These data are well described by the statistical model,
allowing an extraction of the kaon PDFs. We find that both the
Drell-Yan and the $J/\psi$ data favor a harder valence distribution
for strange quark than for up quark in kaon. 
The kaon gluon distribution is further constrained by the 
$J/\psi$ production data. In particular, the momentum fraction 
carried by gluons is found to be similar for pion and kaon. 
\end{abstract}


\end{frontmatter}

The study of proton's partonic substructure has been actively pursued
since the discovery of point-like constituents of the nucleons in
deep-inelastic scattering (DIS) reaction. 
In contrast, the partonic
substructures of pion and kaon, which are the lightest hadrons with the 
dual roles of $q \bar q$ bound states and Goldstone bosons, 
remain poorly studied experimentally~\cite{Holt10}. 
The lack of DIS data on these ephemeral particles represents a major 
limitation for accessing their substructures experimentally. 
Nevertheless, pion-induced 
Drell-Yan process~\cite{Drell70} has provided a first glimpse of the 
valence-quark distribution in pion~\cite{E326DY,E615DY,NA10DY}. 
These data, together with
the pion-induced direct-photon production and the tagged-neutron DIS
data, have led to the extraction of parton 
distributions in pion~\cite{Owens,ABFKW,GRV,SMRS,JAM,xFitter}, although
the sea-quark and gluon distributions remain to be better determined. 
It was pointed out~\cite{Peng17,Chang20,Hsieh21,Chang23}
that pion-induced $J/\psi$ production data could probe the 
gluon as well as the valence-quark distributions in pion. 

Compared to the situation for proton and pion, 
extremely limited information on the partonic structure of kaon
is available experimentally. Based on a
total of $\sim 700$ Drell-Yan events using a $K^-$ beam~\cite{NA3DY}, 
the NA3 collaboration inferred that the $\bar u$ valence quark
in $K^-$ has a softer momentum distribution than that in $\pi^-$.
This reflects the breaking of the SU(3) flavor symmetry, resulting in
a larger fraction of $K^-$'s momentum being carried by the $s$ quarks
than the lighter $\bar u$ quarks. The indication of a
flavor dependence of valence-quark distributions in kaon has generated
much interest, and has inspired  many theoretical 
calculations~\cite{Chang14,Chen16,Hutauruk16,Roberts22,Bednar,Han21}. 
Recent advent in
lattice QCD to calculate the momentum ($x$) dependence of meson 
parton distribution functions (PDFs)~\cite{Ji13,Zhang19,Sufian19,Gao20,
Fan21,Lin21,ETM21,Lin22} as well as the suggestion that
the gluon content of kaon is different from that of pion~\cite{Chen16}, 
has led to new initiatives for collecting additional Drell-Yan data 
with kaon beams~\cite{Adams18}. 

While several sets of pion PDFs have been obtained from global analyses of
existing data, the kaon PDFs have only been extracted based on
a fit to the scarce NA3 $K^-/\pi^-$ Drell-Yan data. It was suggested 
that the more abundant kaon-induced $J/\psi$ 
production data could constrain the kaon 
PDFs~\cite{Peng17}. Indeed, pion-induced $J/\psi$ production data,
together with the Drell-Yan data, were included in a recent global fit 
to extract the pion PDFs in the framework of a statistical 
model~\cite{Bourrely22}.  
This paper reports the extraction of the kaon PDFs
from a global fit to
the kaon-induced Drell-Yan and $J/\psi$ production data in the 
statistical model.

The statistical approach for describing the partonic distributions in hadrons
was initiated over 20 years ago~\cite{Bourrely02}. 
Valuable insights on the flavor and spin structure of the
momentum dependencies of quarks and gluons in the proton have been provided
with the statistical model~\cite{Bourrely05}. 
A salient feature of the statistical model 
is the connection between the valence and the sea quark distributions 
through their helicity-dependent Fermi-Dirac momentum distributions. 
The statistical model 
approach has led to many successful predictions~\cite{Bourrely05}, including 
the flavor asymmetry, $\bar d(x) > \bar u(x)$, for the unpolarized
sea~\cite{Baldit94,Towell01,Dove21} and the inequality, 
$\Delta \bar u(x) > 0 > \Delta \bar d(x)$, for the polarized 
sea~\cite{Adam19}. 

To extract the kaon PDFs in the statistical model, we first define 
the notations of the PDFs of pions and kaons. Imposing the particle-antiparticle 
charge-conjugation (C) symmetry and the isospin symmetry 
for parton distributions~\cite{Londergan10}, we can define the
PDFs of charged pions as follows~\cite{Bourrely21}:
\begin{eqnarray}
U_\pi(x) & \equiv &  u_{\pi^+}(x) = \bar u_{\pi^-}(x) =
\bar d_{\pi^+}(x) = d_{\pi^-}(x)~; \nonumber \\
\bar U_\pi(x) & \equiv & \bar u_{\pi^+}(x) = \bar d_{\pi^-}(x)
= d_{\pi^+}(x) = u_{\pi^-}(x)~; \nonumber \\
S_\pi(x) & \equiv & s_{\pi^+}(x) = \bar s_{\pi^-}(x)
= \bar s_{\pi^+}(x) = s_{\pi^-}(x)~; \nonumber \\
G_\pi(x) & \equiv &  g_{\pi^+}(x) = g_{\pi^-}(x)~.
\label{eq1}
\end{eqnarray}
In the framework of the statistical model, the four pion parton 
distributions are given in the 
following parametric forms~\cite{Bourrely22}:
\begin{eqnarray}
xU_\pi(x) & = & \frac{A_U X_U x^{b_U}}{\exp[(x-X_U)/\bar x] + 1}
+\frac{\tilde A_{U} x^{\tilde b_{U}}}{\exp(x/\bar x) + 1}~; \nonumber \\
x\bar U_\pi(x) & = & \frac{A_U (X_U)^{-1} x^{b_U}}{\exp[(x+X_U)/\bar x]
+ 1} +\frac{\tilde A_{U} x^{\tilde b_{U}}}{\exp(x/\bar x) + 1}~; \nonumber \\
x S_\pi(x) & = & \frac{\tilde A_{U} x^{\tilde b_{U}}}{2[\exp(x/\bar x)
+ 1]}~; \nonumber \\
x G_\pi(x) & = & \frac {A_G x^{b_G}} {\exp(x/\bar x)-1}~,~~~~ 
b_G=1+\tilde b_U~.
\label{eq2}
\end{eqnarray}

The $x$ distributions
for quarks and antiquarks have Fermi-Dirac parametric form, while
gluon has a Bose-Einstein form~\cite{Bourrely02,Bourrely05}.
The two terms in $xU_\pi(x)$ and $x \bar U_\pi (x)$ correspond to
the non-diffractive
and diffractive contributions~\cite{Bourrely02,Bourrely05}.
As shown in~\cite{Bourrely05}, the diffractive term 
is important at the low $x$ region.
A key feature of the statistical model is
that the chemical potential, $X_U$, for the parton distribution 
becomes $-X_U$ for the anti-parton distribution. The parameter $\bar x$
signifies the effective ``temperature".
For the strange-quark distribution, $S_\pi(x)$, the absence of
the valence strange quarks implies that the chemical
potential must vanish for the non-diffractive term.  
The assumption that $S_\pi(x)$ equals half of the diffractive part
of  $\bar U_\pi(x)$ reflects the heavier strange quark mass.
The expression $b_G = 1+\tilde b_U$ ensures that $G(x)$ has an
identical $x$ dependence as the diffractive part of the quark distributions 
when $x \to 0$.

For charged kaons, as $K^+$ and $K^-$ belong to different isospin
multiplets, only the charge-conjugation symmetry is applicable.
The kaon PDFs are then defined as: 
\begin{eqnarray}
U_K(x) & \equiv & u_{K^+}(x) = \bar u_{K^-}(x)~; \nonumber \\
S_K(x) & \equiv & \bar s_{K^+}(x) = s_{K^-}(x)~; \nonumber \\
D_K(x) & \equiv & d_{K^+}(x) = \bar d_{K^-}(x)~; \nonumber \\
G_K(x) & \equiv &  g_{K^+}(x) = g_{K^-}(x)~,
\label{eq4}
\end{eqnarray}
with analogous expressions for $\bar U_K(x), \bar S_K(x)$ and
$\bar D_K(x)$. 
The kaon PDFs are constrained by the valence-quark 
and the momentum sum rule:
\begin{eqnarray}
 & \int_0^1 [U_K(x) - \bar U_K(x)]~dx =1~, \nonumber \\
 & \int_0^1 [S_K(x) - \bar S_K(x)]~dx =1~, \nonumber \\
 & \int_0^1 x [U_K(x) + \bar U_K(x) +S_K(x) +\bar S_K(x) + \nonumber \\
 & 2 D_K(x) + G_K(x)]~dx = 1~.
\label{eq5}
\end{eqnarray}
The kaon parton distributions in the 
statistical model are expressed in the following parametric forms:
\begin{eqnarray}
xU_K(x) & = & \frac{A_{UK} X_{UK} x^{b_{UK}}}{\exp[(x-X_{UK})/\bar x] + 1}
+\frac{\tilde A_{UK} x^{\tilde b_{UK}}}{\exp(x/\bar x) + 1}~; \nonumber \\
x\bar U_K(x) & = & \frac{A_{UK} 
(X_{UK})^{-1} x^{b_{UK}}}{\exp[(x+X_{UK})/\bar x]
+ 1} +\frac{\tilde A_{UK} x^{\tilde b_{UK}}}{\exp(x/\bar x) + 1}~; \nonumber \\
xS_K(x) & = & \frac{A_{SK} X_{SK} x^{b_{SK}}}{\exp[(x-X_{SK})/\bar x] + 1}
+\frac{\tilde A_{UK} x^{\tilde b_{UK}}}{2[\exp(x/\bar x) + 1]}~; 
\nonumber \\
x\bar S_K(x) & = & \frac{A_{SK} (X_{SK})^{-1} 
x^{b_{SK}}}{\exp[(x+X_{SK})/\bar x]+ 1} 
+\frac{\tilde A_{UK} x^{\tilde b_{UK}}}{2[\exp(x/\bar x) + 1]}~; 
\nonumber \\
x D_K(x) & = & x \bar D_K(x) = \frac{\tilde A_{UK} 
x^{\tilde b_{UK}}}{(\exp(x/\bar x) + 1)}~; \nonumber \\
x G_K(x) & = & \frac {A_{GK} x^{b_{GK}}} {\exp(x/\bar x)-1}~,~~~~
b_{GK}=1+\tilde b_{UK}~.
\label{eq6}
\end{eqnarray}

To obtain the parameters for kaon PDFs, we have
fitted both the Drell-Yan and the $J/\psi$ production data.
The only available Drell-Yan data with kaon beam are from the NA3
collaboration, which performed a 
simultaneous measurement of $K^- +$ Pt $\to \mu^+ \mu^- + X$ and
$\pi^- +$ Pt $\to \mu^+ \mu^- + X$ using a 150 GeV beam on a platinum
target~\cite{NA3DY}. 
Figure 1 shows the $K^- / \pi^-$ Drell-Yan cross section ratios from NA3
as a function of
$x_1$, the fraction of the beam momentum carried by
the interacting parton. Figure 1 shows that the ratio $R$ falls below unity
at large $x_1$ ($x_1 > 0.65$). Since the Drell-Yan cross sections with 
$\pi^-$ and $K^-$ beams at large $x_1$ are dominated by the term containing 
the product of $\bar u_M(x_1)$ in the meson $M$
and $u_A(x_2)$ in the nucleus $A$, the fall-off in $R$ at large $x_1$  
indicates that $U_K(x)$
is softer than $U_\pi (x)$~\cite{NA3DY}.

\begin{figure}[tb]
\begin{center}
\includegraphics[height=5.3cm, width=5.5cm]{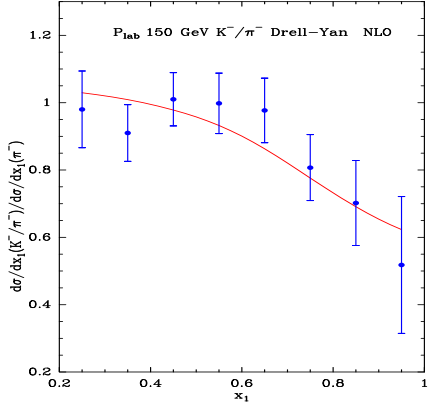}
\end{center}
\caption{\baselineskip 1pt
$K^-/\pi^-$ Drell-Yan cross section ratio data from the NA3 experiment 
at beam momentum of 150 GeV on a platinum target~\cite{NA3DY}.
The data are compared with NLO calculation using the kaon PDFs obtained
in this analysis and the pion PDFs obtained in a recent 
analysis~\cite{Bourrely22} using
the statistical model.}
\label{fi1}
\end{figure}

We have performed a next-to-leading-order (NLO)
QCD calculation to fit the NA3 $K^-/\pi^-$ Drell-Yan data. 
Detailed expressions for the NLO Drell-Yan cross
sections can be found in~\cite{Bourrely19}. The nucleon PDFs
used in the calculation were taken from the BS15 PDFs~\cite{Bourrely15},
obtained from a global fit to existing data in the
framework of the statistical model. The QCD evolution was performed using
the HOPPET program~\cite{Salam2009}, and the CERN MINUIT
program~\cite{James1994} was utilized for the $\chi^2$
minimization. Since the NA3 Drell-Yan data were collected using
nuclear targets (platinum), we take into account the
nuclear modification of the nucleon PDFs, described in~\cite{Bourrely21}.

\begin{figure}[tb]
\centering
\subfigure[]
{\includegraphics[width=3.9cm,  height=3.6cm]{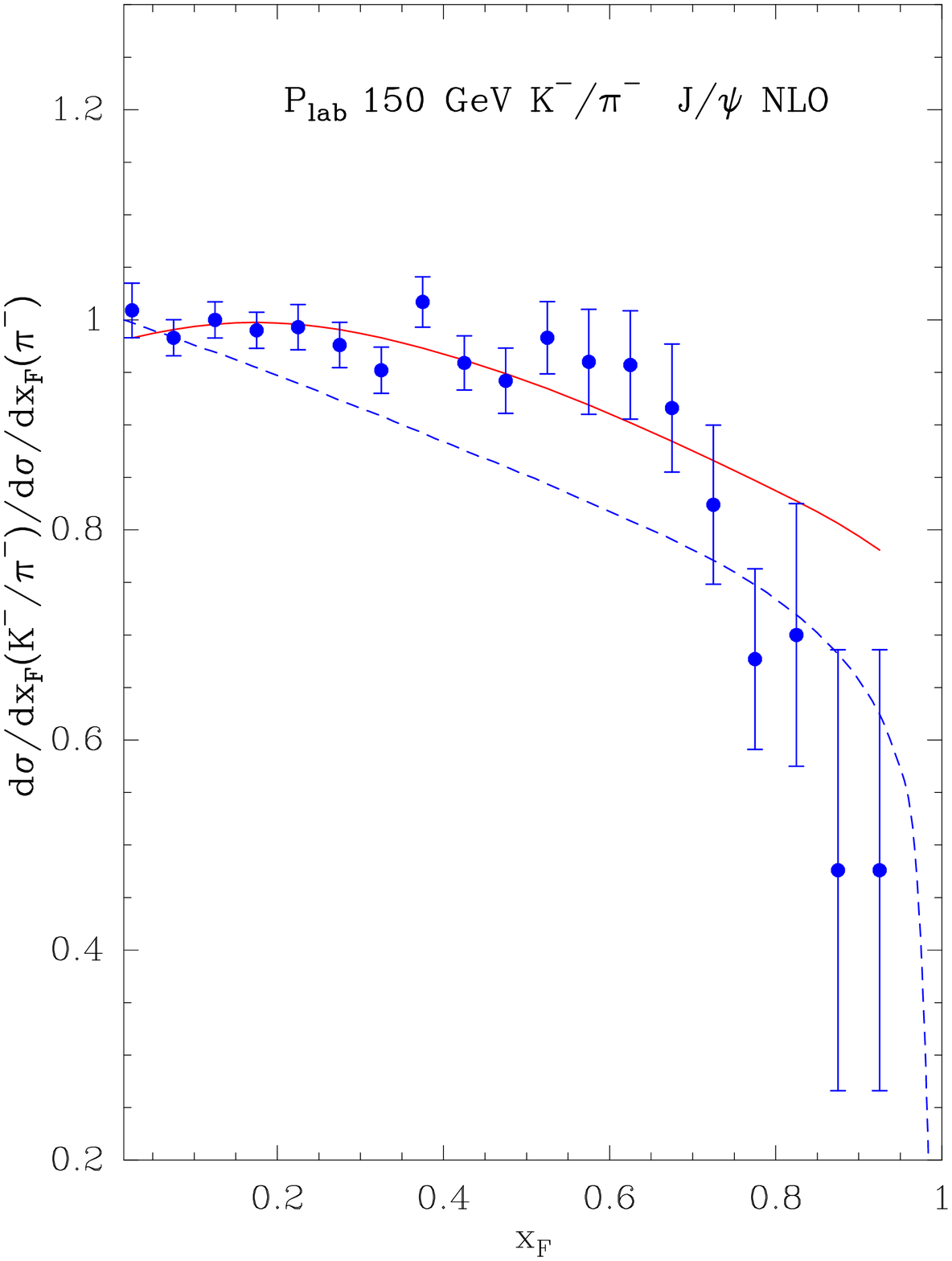}}
\subfigure[]
{\includegraphics[width=3.9cm,  height=3.6cm]{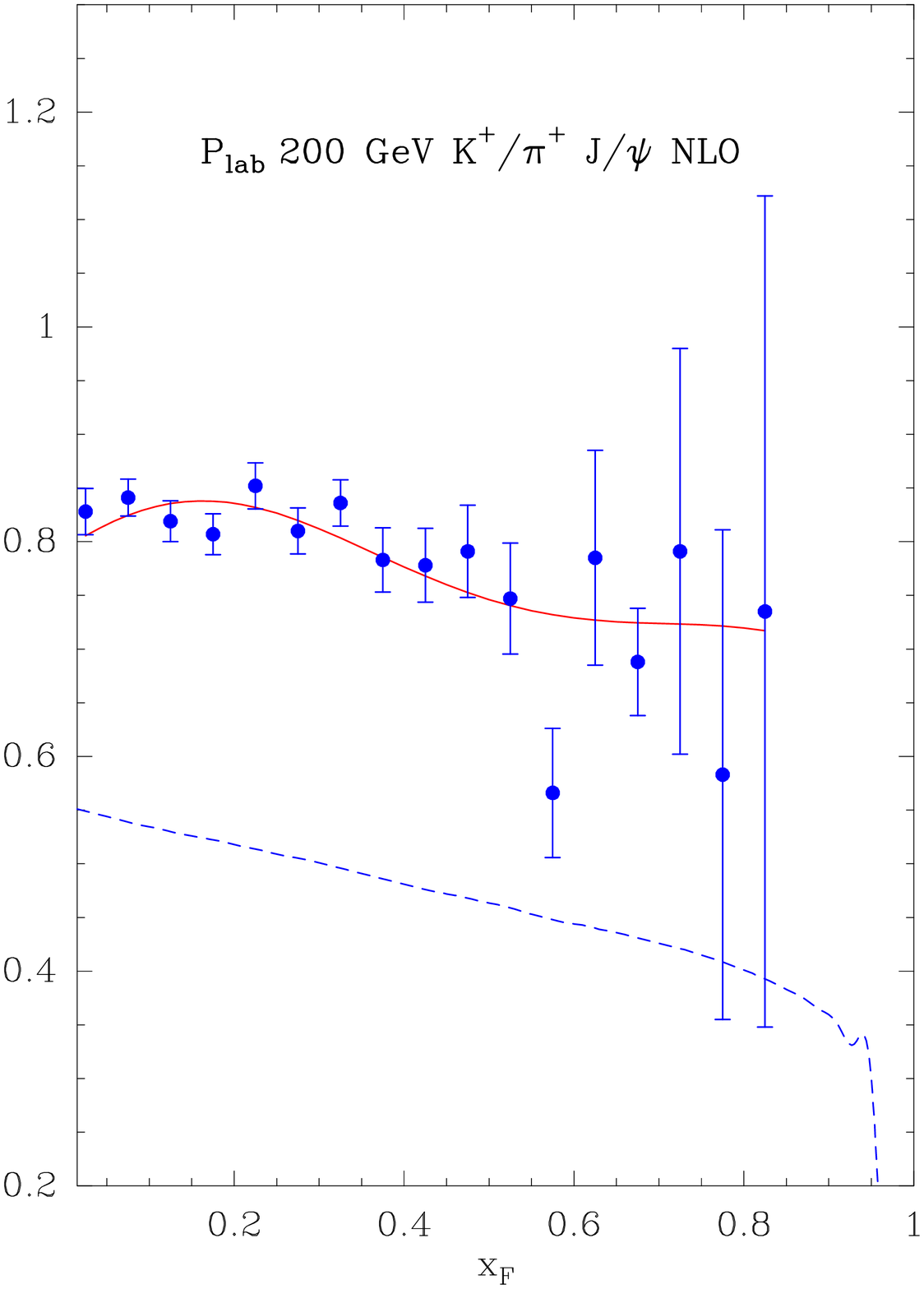}}\\
\subfigure[]
{\includegraphics[width=3.9cm,  height=3.6cm]{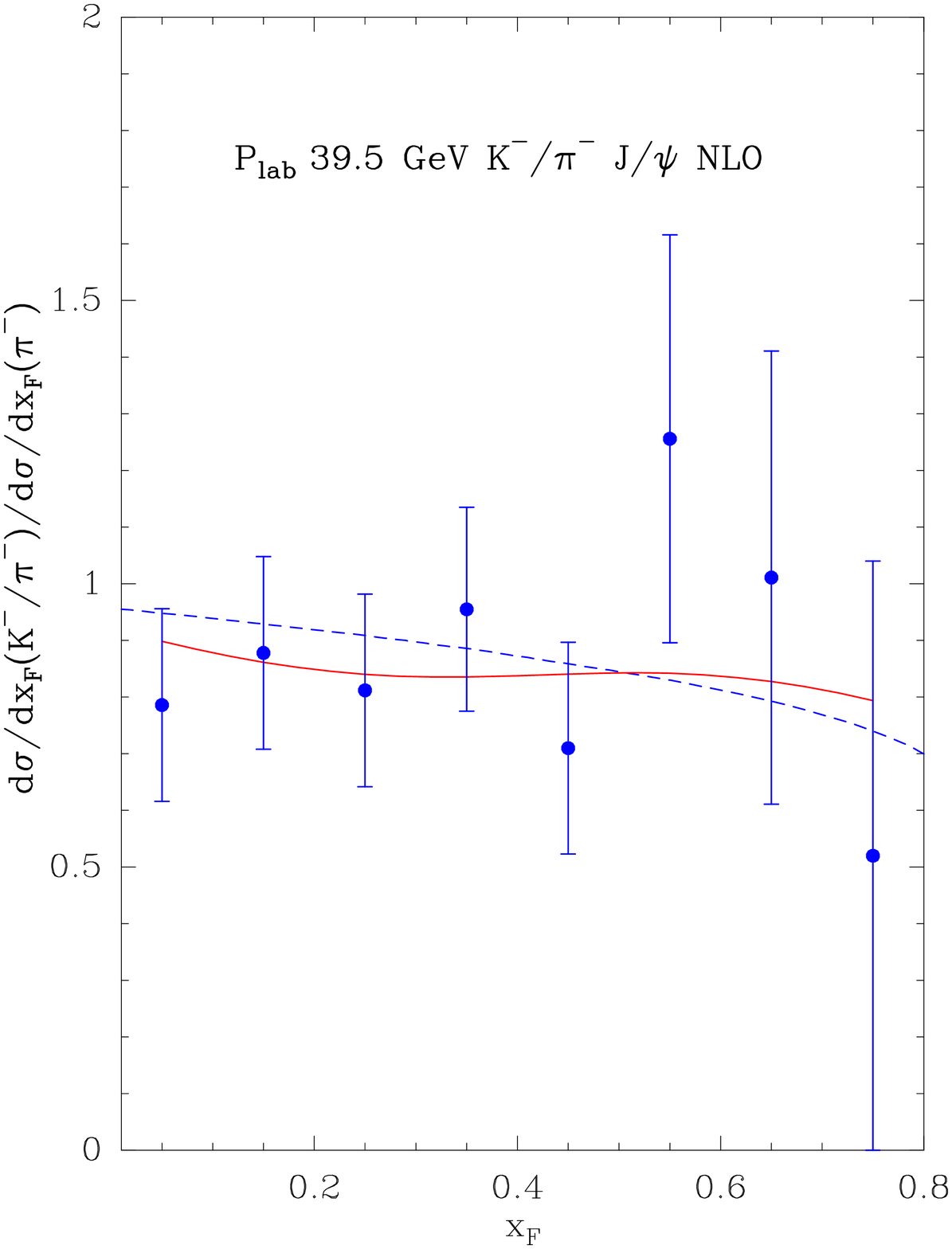}}
\subfigure[]
{\includegraphics[width=3.9cm,  height=3.6cm]{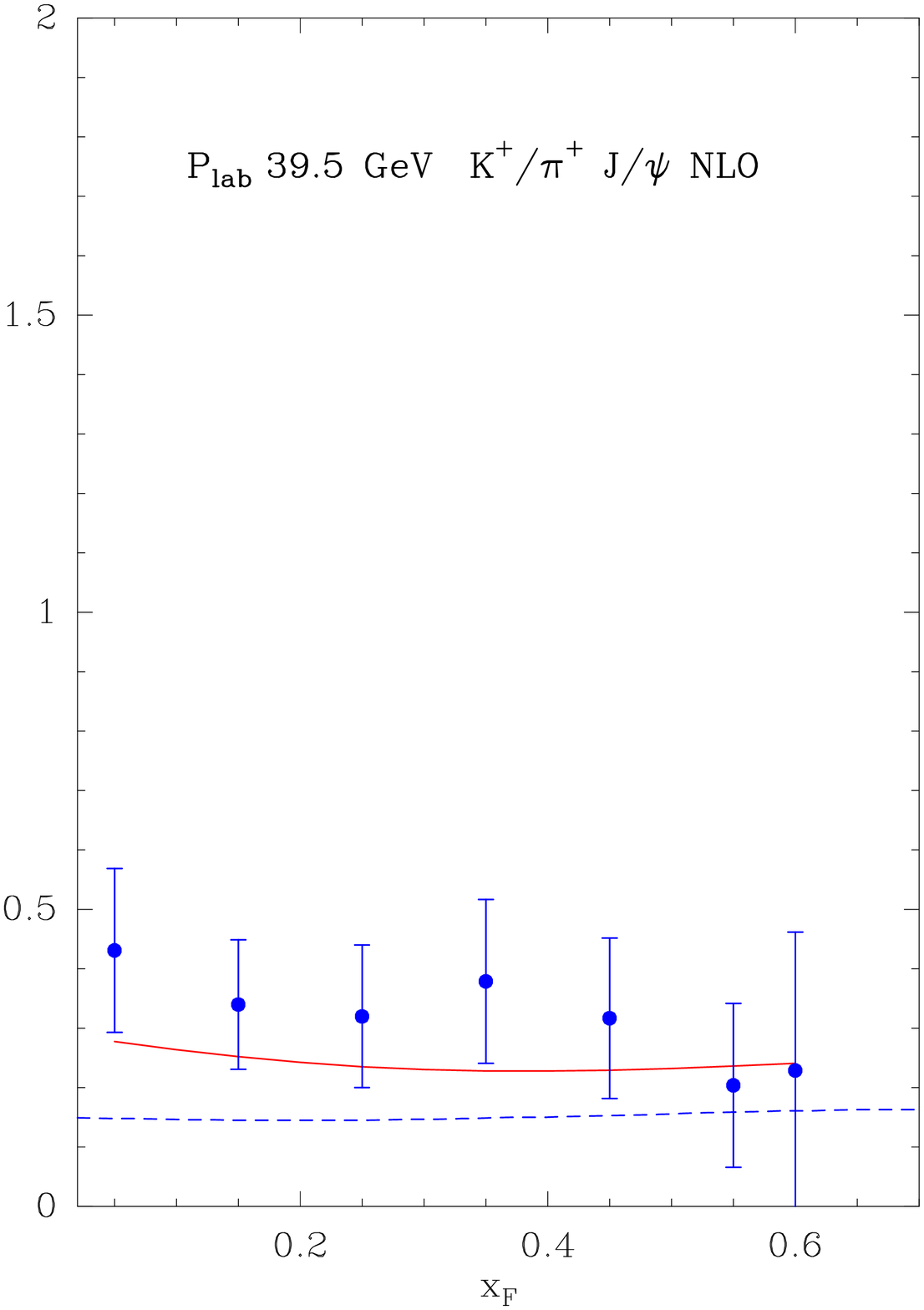}}
\caption{ $J/\psi$ cross section ratios data on platinum targets for
(a) $K^-/\pi^-$ at 150 GeV~\cite{NA3JPSI},
(b) $K^+/\pi^+$ at 200 GeV~\cite{NA3JPSI},
(c) $K^-/\pi^-$ at 39.5 GeV~\cite{WA39}, and
(d) $K^+/\pi^+$ at 39.5 GeV~\cite{WA39}.
The solid curves are NRQCD calculation using the kaon and pion
PDFs obtained in the statistical model, while the dashed curves correspond
to NRQCD calculation using the meson PDFs from Ref.~\cite{Han21}.
}
\label{fig2}
\end{figure}

The NA3 collaboration also reported the measurement
of $K^- / \pi^-$ ratio versus $x_F$ ($x$-Feynman) 
for $J/\psi$ production at 150 GeV 
on a platinum target, shown in Fig. 2(a)~\cite{NA3JPSI}.
While the 
$K^- / \pi^-$ ratio is relatively flat for the region $0 < x_F < 0.6$,
it starts to drop noticeably when $x_F$ further increases.
A comparison between Fig. 1 and Fig. 2(a) shows a striking similarity.
Since the fall-off at large $x_1$ in the $K^- / \pi^-$ Drell-Yan
cross section ratio is described
as a soft $U_K(x)$ distribution,
it is conceivable that the pronounced drop of the $K^- / \pi^-$ ratio
at large $x_F$ in Fig. 2(a) has a similar origin.

The NA3 collaboration has also measured the $K^+ / \pi^+$ ratios for 
$J/\psi$ production at 200 GeV 
on a platinum target as shown in Fig. 2(b)~\cite{NA3JPSI}.
A significant difference between the $K^+ / \pi^+$ and the 
$K^- / \pi^-$ ratios is observed. While there is a pronounced drop
of the $K^- / \pi^-$ ratio at forward $x_F$, no such drop is present
for $K^+ / \pi^+$. Moreover, Figures 2(a) and 2(b) show that
the $K^+/\pi^+$ ratios are $\sim$ 20\% lower than 
for the $K^-/\pi^-$ ratio.

The only other $J/\psi$ production data with kaon beams were obtained
by the WA39 collaboration using 39.5 GeV beam on a 
tungsten target~\cite{WA39}.
Both the $K^- / \pi^-$ and the $K^+ / \pi^+$ $J/\psi$ cross section
ratios were measured, as shown in Fig. 2(c) and Fig. 2(d). 
The $K^+/\pi^+$ ratios at 39.5 GeV are notably lower than that at 200 GeV.
These $K/\pi$ $J/\psi$ production data, 
together with the $K^-/\pi^-$ Drell-Yan data, are utilized in this analysis 
for the first determination of the kaon PDFs. The striking energy dependence 
of the $K^+/\pi^+$ $J/\psi$ ratios, as well as the difference between the 
$K^-/\pi^-$ and $K^+/\pi^+$ $J/\psi$ ratios, suggest the possibility of 
flavor separation, namely, to
distinguish the quark and gluon distributions in koan, as discussed later. 

For the calculation of the $J/\psi$ production cross section, we adopt the
non-relativistic QCD (NRQCD)~\cite{NRQCD} approach. The NRQCD
framework, which was also used in the recent analysis of the
pion-induced $J/\psi$ production data to extract the pion
PDFs~\cite{Bourrely22}, is
based on the factorization of the heavy-quark $Q\bar{Q}$ pair production
and its subsequent hadronization. The production of the $Q\bar{Q}$
pair involves short-distance partonic interaction,
calculated using perturbative QCD. The subprocesses include the gluon-gluon
fusion, quark-antiquark annihilation, and quark-gluon interaction. 

In NRQCD the probability of
a $Q\bar{Q}$ pair hadronizing into a quarkonium bound state
is described by the long-distance matrix
elements (LDMEs). The LDMEs, assumed to be universal and independent of the
beam type, are determined from the experimental 
data~\cite{Beneke:1996tk}.
The LDMEs used in the NRQCD calculation were taken from a recent
study~\cite{Hsieh21}, which extracts these matrix elements by
performing a global fit to the 
$J/\psi$ production cross sections induced by proton and pion beams at
fixed-target energies. Several sets of the LDMEs were obtained in this
work~\cite{Hsieh21} and we have selected the ``Fit-2" solution.  We
found that the results of the present analysis are insensitive to
the choice of the specific LDME set.  

\begin{figure}[tb]
\centering
\subfigure[]
{\includegraphics[height=3.0cm, width=5.0cm]{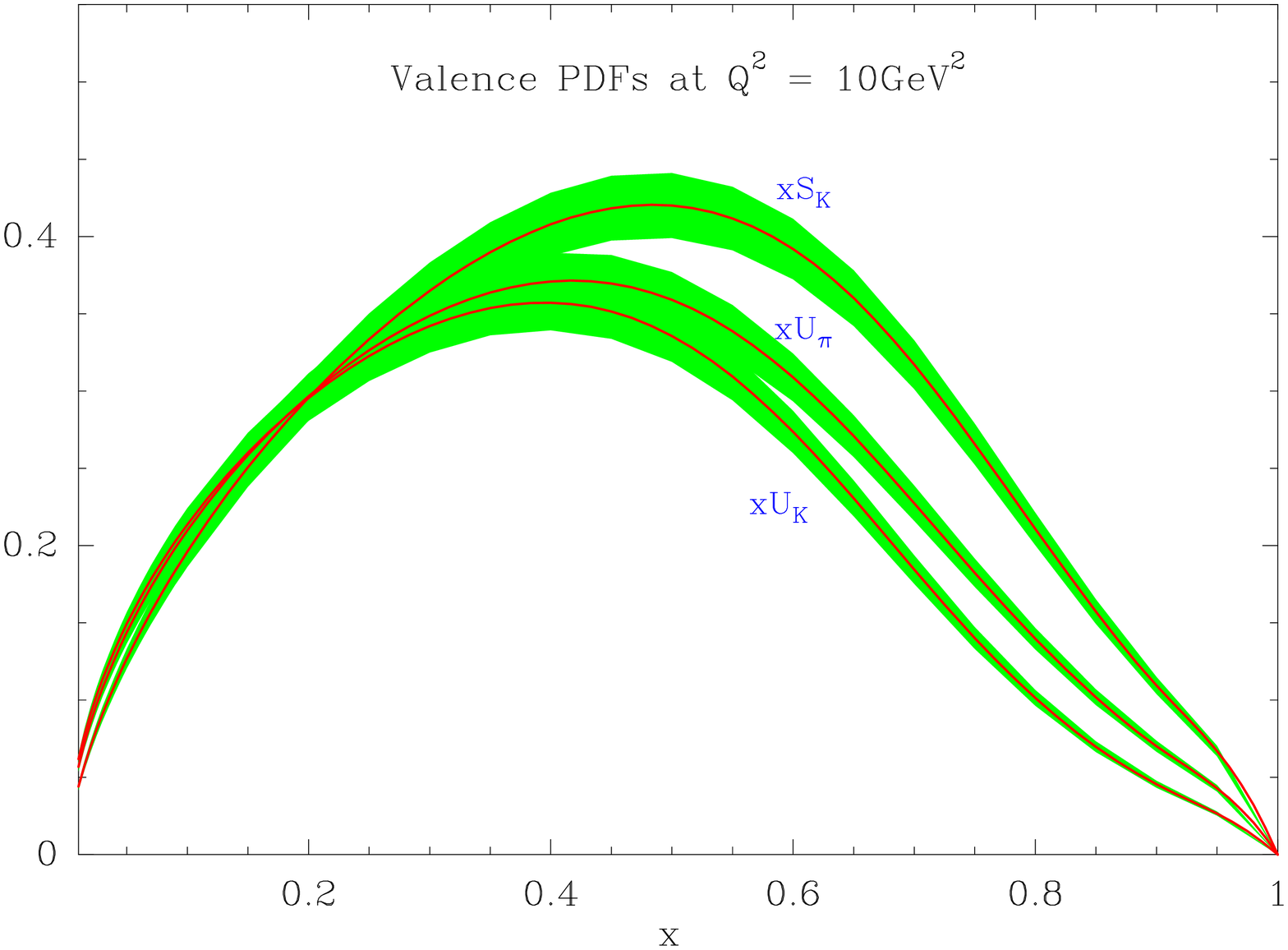}}
\subfigure[]
{\includegraphics[height=3.0cm, width=5.0cm]{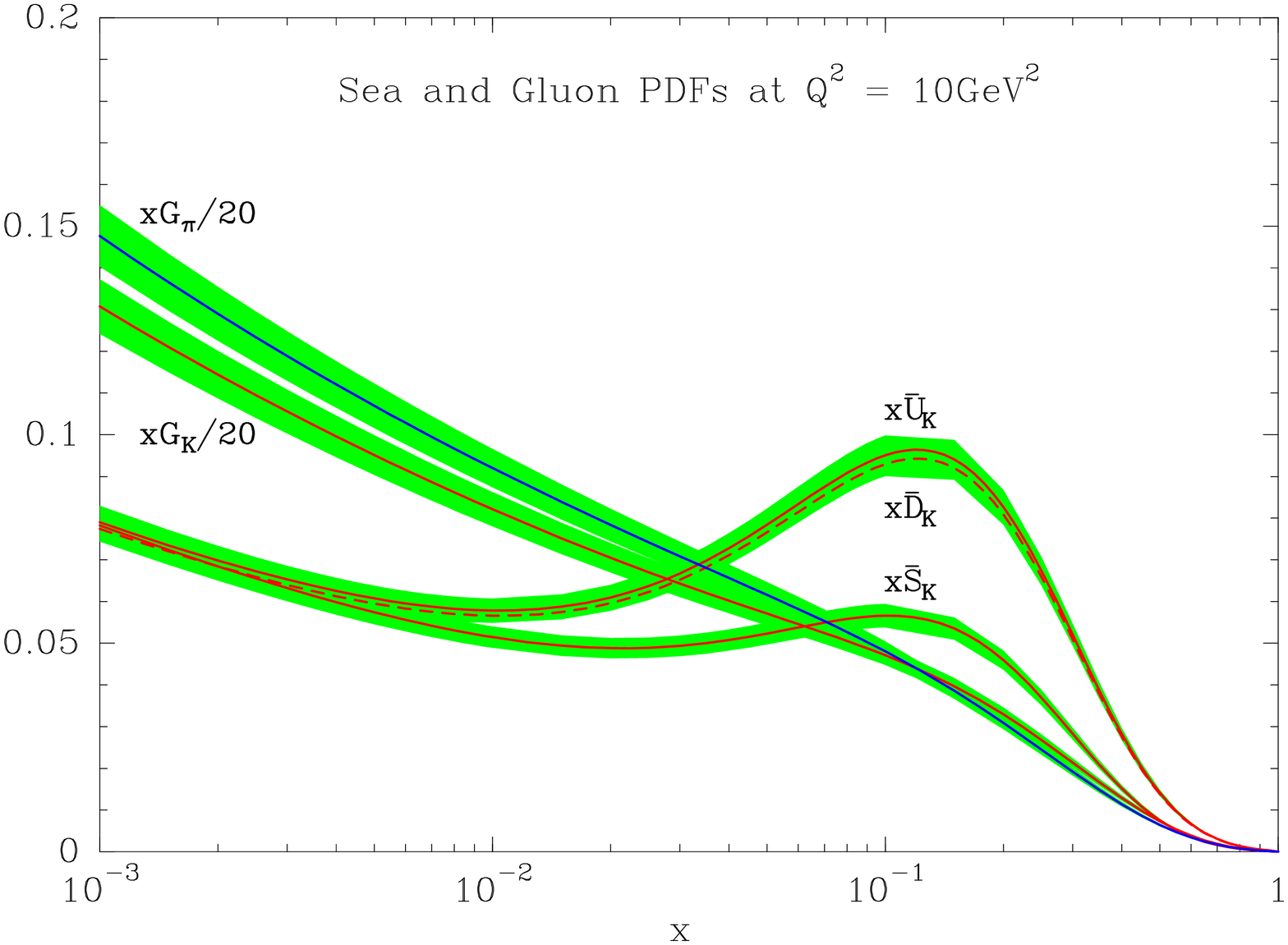}}
\caption{\baselineskip 1pt
Kaon PDFs obtained from a fit to the $K/\pi$ ratios from Drell-Yan and 
$J/\psi$ production experiments in the statistical model. 
(a): Valence quark distributions and (b): Sea-quark and gluon
distributions. The pion quark and
gluon distributions are also shown for comparison.
}
\label{fi3}
\end{figure}

Since the available Drell-Yan and $J/\psi$ data used in this analysis are
all in the form of the $K/\pi$ cross section ratios, both the pion and the
kaon PDFs are needed for the calculation. As the pion PDFs were already 
extracted in the framework of the statistical model, we
fix the pion PDFs according to the results obtained from this recent 
study~\cite{Bourrely22} while allowing the kaon PDFs to vary. The best-fit
values for the various parameters in the statistical model are
obtained for kaon. Table I lists the number of data points and the 
values of $\chi^2$ for the best fit to the various data. 
In the global fit, the 
normalizations for various data sets
are allowed to vary, i.e., the result of the
calculated $K/\pi$ ratio is multiplied by a K factor when compared 
with the data. We find that the K factors for the fit to Drell-Yan 
and $J/\psi$ data are very close to 1 for negative mesons, and
the K factors for positive mesons are within 18\% of unity, 
consistent with the normalization uncertainties of the experiments.  
\begin{table}[htbp]
\caption {Values of the K factor $\chi^2$ for each
data set obtained from a global fit.
P is the beam momentum, K the 
factor to be multiplied to the calculated Drell-Yan
and $J/\psi$ cross section ratios,
and $ndp$ 
the number of data points.
}
\begin{center}
\scalebox{0.9}{
\begin{tabular}{ c c c c c c}
\hline
\hline
Experiment   &P(GeV) &K  & $ndp$ & $\chi^2$ & $\chi^2/ndp$ \\
\hline\raisebox{0pt}[12pt][6pt]

$K^-/\pi^-$ DY NA3   & 150    & 1.052   & 8   & 4.05 & 0.51 \\[4pt]
$K^-/\pi^- J/\psi$ WA39    & 39.5    & 0.98   & 8  & 3.7 & 0.45 \\[4pt]
$K^-/\pi^- J/\psi$ NA3     & 150    & 1.028   & 19  & 21.2 & 1.12  \\[4pt]
$K^+/\pi^+ J/\psi$ WA39    & 39.5    & 1.15   & 7  & 4.0  &  0.58\\[4pt]
$K^+/\pi^+ J/\psi$ NA3 & 200   &  1.18   & 17  & 11.0  & 0.65 \\[4pt]
\hline\raisebox{0pt}[12pt][6pt]
Total                        & &          & 59  &  43.95 & 0.74\\[4pt]
\hline
\hline
\end{tabular}
\label{table1}
}
\end{center}
\end{table}

The small $\chi^2/ndp$ values listed in Table I show that a satisfactory
description of the Drell-Yan and $J/\psi$ $K/\pi$ data can be achieved in the
statistical model. This is also shown in Fig. 1 and Fig. 2,
where the $K/\pi$ data for the Drell-Yan and the $J/\psi$ production 
are compared with the calculations. 
To illustrate the importance of including the $J/\psi$ data in extracting
the kaon PDFs, the dashed curves in Fig. 2 show the NRQCD calculation using
the kaon and pion PDFs obtained in the maximum 
entropy approach~\cite{Han21}. The poor
agreement with the data suggests that the sparse Drell-Yan data alone are
not sufficient to determine the kaon PDFs. 

The best-fit parameters of the kaon PDFs, obtained at an initial
scale $Q^2_0 = 1$ GeV$^2$, are:

\begin{align}
A_{UK}  &=  1.12 \pm 0.05 & b_{UK} & = 0.602 \pm 0.017  \nonumber \\
X_{UK}  &=  0.688 \pm 0.01   & \bar x & = 0.109\pm 0.001  \nonumber \\
\tilde A_{UK}  &= 4.83 \pm 0.74 & \tilde b_{UK} & = 1.248 \pm 0.05 \nonumber \\
A_{SK}  &=  1.14 \pm 0.02 & b_{SK} & = 0.73 \pm 0.009  \nonumber \\
X_{SK}  &=  0.784 \pm 0.01   & A_{GK}  &=  108.97 \pm 1.0~.
\label{eq14}
\end{align}

\begin{table}[htbp]
\caption {Momentum fractions of valence quarks, 
sea quarks, and gluons for $\pi^-$ and $K^-$ at the 
scale $Q^2$= 10 GeV$^2$ obtained in the statistical model.}
\begin{center}
\scalebox{0.8}{
\begin{tabular}{|c|c|c|c|c|c|}
\hline
\hline
  & $ u$ Valence & $d$ Valence & $s$ Valence & all Sea & Gluon \\
\hline
$\pi^-$ & \small{$0.242 \pm 0.004$} & \small{$0.242 \pm 0.004$} 
& $-$ & \small{$0.188 \pm 0.004$} & \small{$ 0.326 \pm 0.015$} \\
$K^-$ & \small{$0.220 \pm 0.002$} & $-$ & \small{$0.276 \pm 0.001$} 
& \small{$0.162 \pm 0.006$}& 
\small{$0.331 \pm 0.018$} \\
\hline
\hline
\end{tabular}
}
\end{center}
\label{tab:xmoment}
\end{table}

\noindent The uncertainties of the parameters in Eq. (6) are
statistical only. The $K/\pi$ cross section ratios are not
sensitive to the uncertainty of the nuclear PDFs. To evaluate the
impact of the uncertainty of the pion PDFs on the extracted kaon PDFs,
a future analysis to fit simultaneously the pion and kaon data is
anticipated. The temperature, $\bar x = 0.109$, found for kaon is very close
to that obtained for pion, $\bar x = 0.119$~\cite{Bourrely22}, indicating a
common feature for the statistical description for pion and kaon.
On the other hand, the chemical potential for the valence quark of 
pion, $X_U =0.72$, 
is between the corresponding chemical potentials of $X_{UK} =0.688$
and $X_{SK}=0.784$ for kaon. The global fit to the 
existing meson-induced Drell-Yan and $J/\psi$ production data 
in the statistical model naturally leads to the result,
$X_{SK} > X_U > X_{UK}$. The smaller chemical potential for $\bar u$ in 
$K^-$ than $\bar u$ in $\pi^-$ accounts for a softer $x$ distribution 
for $U_K(x)$
than $U_\pi (x)$, resulting in the drop of the $K^-/\pi^-$ ratios at the 
large $x_1 (x_F)$ region for both the Drell-Yan and the $J/\psi$ production 
data. 
Figure 3(a) displays the valence quark distributions
at $Q^2 = 10 $  GeV$^2$ for kaon and pion obtained in the statistical mode
analysis. The hierarchy that 
$S_K(x)$ is harder than $U_\pi(x)$, together with $U_\pi(x)$ being harder than
$U_K(x)$, is preserved at the $J/\psi$ production scale of $Q^2 = 10$ GeV$^2$.
The sea-quark and gluon distributions are shown in Fig. 3(b). The 
gluon distributions are found to be very similar for pion and kaon.

To obtain some insight on how the $K^-/\pi^-$ and $K^+/\pi^+$ ratios for 
$J/\psi$ production can constrain the kaon PDFs, we have examined 
the decomposition
of the $J/\psi$ production cross sections for kaon beam into the $q \bar q$
annihilation and the $gg$ fusion processes. 
For $K^-$ beam at 39.5 GeV, the $q \bar q$ annihilation is the dominant
subprocess, while the $gg$ fusion becomes more important at
the higher energy of 150 GeV. Consequently, the combination of the
$J/\psi$ production data at the 39.5 and 150 GeV could constrain
both the valence-quark and the gluon contents in kaon. For the $K^+$ beam,
which contains the $u$ and $\bar s$ valence quarks, the $q \bar q$
annihilation process is suppressed since a sea quark in the nucleon is
required. Hence the $gg$ fusion process is important
for $K^+$ not only at 200 GeV but also at the lower beam energy of 39.5 GeV. 
Thus, the $K^+/\pi^+$ data at both beam energies are sensitive to
the gluon distribution of kaon.

Table II lists the momentum fractions carried by the valence
quarks, sea quarks, and
gluons in $K^-$ obtained in this work
at $Q^2 = 10$ GeV$^2$. The corresponding
momentum fractions for $\pi^-$ obtained in the
statistical model analysis~\cite{Bourrely22} are also shown 
for comparison. As discussed above, the softer $x$ distribution for
$\bar u$ valence quark in $K^-$ leads to a smaller $\bar u$ momentum 
fraction in $K^-$ than in $\pi^-$. Table II also shows that the 
momentum fraction
carried by gluons in kaon is comparable to that in pion. This finding
is at variance with the prediction of Ref.~\cite{Chen16}, but consistent
with the expectation of Ref.~\cite{Roberts22}. As the gluon contents
in the pion and kaon are better constrained when the  $J/\psi$ production
data are included in the global analysis, this work provides the first
evidence that the momentum fractions taken by the gluons are comparable
for the pion and kaon.

In summary, we have performed an extraction of
kaon's PDFs in the statistical
model from a global fit to existing kaon-induced 
Drell-Yan and $J/\psi$ production data.
These data are well described by the statistical model approach.
The inclusion of the $K/\pi$ data for $J/\psi$ production  
provides additional constraints on the valence as well as the
gluon distributions of kaon.
Both the Drell-Yan and
$J/\psi$ production data favor a harder valence distribution for
strange quark than for up quark in kaon. 
A simultaneous fit to the $K^-/\pi^-$ and $K^+/\pi^+$ $J/\psi$ data
allows a sensitive
determination of the gluon distribution in the kaon, and the momentum
fractions carried by gluon are found to be similar for pion and kaon.
New kaon-induced Drell-Yan and $J/\psi$ production data anticipated from
AMBER~\cite{Adams18} would provide
further constraints on the kaon PDFs. 

This work was supported in part by the U.S. National
Science Foundation Grant No. PHY-1812377 and the National Science 
and Technolgy Council of Taiwan (R. O. C.).

\end{document}